\documentclass[12pt,letterpaper]{article}

\usepackage{datetime}
\usepackage{amsmath,amssymb,array,calc,rotating,epsfig,psfrag, amscd}
\usepackage{color}
\usepackage{verbatim}
\usepackage[colorlinks=false,
    linktoc=all, 
    pdfstartview=FitV,
    bookmarksopen=true]{hyperref}
\usepackage[left=2cm,top=1cm,right=3cm,nohead]{geometry}
\usepackage[english]{babel}

\renewcommand{\d}{\textrm{d}}

\newcommand{\bea}{\begin{eqnarray}}
\newcommand{\eea}{\end{eqnarray}}
\newcommand{\be}{\begin{equation}}
\newcommand{\ee}{\end{equation}}

\definecolor{cardinal}{rgb}{0.6,0,0}
\definecolor{darkgreen}{rgb}{0,0.5,0}
\definecolor{golden}{rgb}{0.92, 0.7, 0}
\definecolor{midnight}{rgb}{0, 0, 0.5}
\definecolor{darkblue}{rgb}{0.2, 0, 0.8}

\newcommand{\beq}{\begin{equation}\begin{aligned}}
\newcommand{\eeq}{\end{aligned}\end{equation}}

\numberwithin{equation}{section}

\begin{document}
\begin{flushright}
\small
UUITP-14/14
\normalsize
\end{flushright}

\thispagestyle{empty}

\begin{center}
\baselineskip=13pt {\LARGE \bf{Fatal attraction:\\ \vspace{0.4cm} more on decaying anti-branes}}
 \vskip1.5cm 
Ulf H. Danielsson$^a$
 and Thomas Van Riet$^c$  \\
\vskip0.5cm
 \textit{$^a$ Institutionen f{\"o}r fysik och astronomi,\\ Uppsala Universitet, Uppsala, Sweden}\\
\href{mailto:ulf.danielsson@physics.uu.se}{ulf.danielsson @ physics.uu.se}

\vskip0.5cm
 \textit{$^c$Instituut voor Theoretische Fysica, K.U. Leuven,\\
  Celestijnenlaan 200D, B-3001 Leuven, Belgium   }\\
\href{mailto:thomas.vanriet @ fys.kuleuven.be}{thomas.vanriet @ fys.kuleuven.be}
\vskip3.5cm
\end{center}

\begin{abstract}
\noindent
We elaborate on the decay of branes inside throat geometries that are supported by flux carrying charges opposite to the brane. Our main point is that such backgrounds necessarily have a local, possibly diverging, pile up of brane-charges dissolved in flux around the anti-brane due to the (fatal) attraction of the flux towards the brane.  We explain that this causes enhanced brane-flux annihilation and is in tension with the idea that anti-branes can be used to construct meta-stable vacua. We argue that stable configurations -- if they at all exist -- are not obtainable within SUGR.  
The problem we point out is already present when the back-reaction is confined in the IR and the associated uplift energy small. Our results are valid in the regime that is complementary to a recent analysis of Bena et.~al.
\end{abstract}

\clearpage

\section{Introduction}\label{sec:intro}
An elegant mechanism for breaking supersymmetry in flux  backgrounds involves the addition of D$p$ branes that are not mutually BPS with the fluxes \cite{Maldacena:2001pb, Kachru:2002gs}, from here onwards referred to as `anti-branes'. The SUSY breaking scale can be tuned small in case the space-time contains a region of large warping at the position of the anti-brane. This occurs dynamically since the anti-brane minimizes its energy in that region. String compactifications of type IIB string theory can naturally contain throat regions of high warping \cite{Dasgupta:1999ss, Giddings:2001yu} and therefore anti-branes became an essential ingredient for de Sitter model building \cite{Kachru:2003aw} and inflation model building \cite{Kachru:2003sx, DeWolfe:2004qx} in string theory. 

As always, once SUSY is broken, one needs to be cautious when it comes to the meta-stability of the vacuum. There exist various decay channels towards the SUSY vacuum and each of them need to be classically forbidden in order to claim a meta-stable vacuum. In a compactified model the stability with respect to the compactification moduli was verified for a simple toy model in \cite{Kachru:2003aw}. But the more subtle stability is in the open string sector where brane-flux annihilation can take place \cite{Kachru:2002gs}. In short, brane-flux annihilation is the process in which fluxes can materialise into actual branes that consequently annihilate with the anti-branes in the throat. This is possible from the point of view of charge conservation because the fluxes carry brane charges themselves. For instance for the case we study in this paper we have 3-form fluxes $F_3$ and $H_3$ that source the 5-form field strength as a smooth charge density
\begin{equation}
\d F_5 = H_3\wedge F_3 + \delta\epsilon_6\,,
\end{equation}
whereas the delta function symbolises a localised 3-brane source ($\epsilon_6$ is the volume element on the internal manifold). Whenever the $\delta$ source term has an opposite orientation with respect to the term involving the 3-form fluxes, SUSY is broken since there are two charge densities of opposite sign coexisting\footnote{This is not strictly true when the vacuum is AdS. See for instance \cite{Apruzzi:2013yva}.}. This would be perturbatively unstable in case both charge densities were consistent of brane sources. But as one term now consists of fluxes one can create a meta-stable state because a non-perturbative effect is required that materializes branes out of the flux densities, for an annihilation to occur. More precisely,
it was verified in \cite{Kachru:2002gs} that in the non-compact Klebanov-Strassler (KS) throat the anti-D3 branes have a classical barrier against brane-flux annihilation if their charge is small enough with respect to the background flux and if it can be treated as a probe. The latter two assumptions naturally fit together. In compactified models the situation can be different,  but the main feature of meta-stability is expected to survive \cite{Frey:2003dm, Brown:2009yb}.

Despite the naturalness and elegance of this idea there are possible caveats to this mechanism. The problematic assumption is the validity of the probe computation as carried out in \cite{Kachru:2002gs}. Brane-flux annihilation proceeds via the Myers effect \cite{Myers:1999ps} in which the anti-D3 brane puffs into a spherical NS5-brane. This can be analysed from the point of view of the non-Abelian anti-D3 action or the Abelian NS5 action. Both actions are probe actions. But in the probe limit one sends the anti-D3 charge to zero and one looses control over the supergravity approximation since the size of the NS5 sphere becomes of stringy scale. Secondly,  the NS5-branes are strongly coupled and it is not clear whether the classical action can be trusted.  So it is difficult to make it fully sound that the probe computation as such is valid. 

\begin{figure}[h!]
\centering
\includegraphics[width=.35\textwidth]{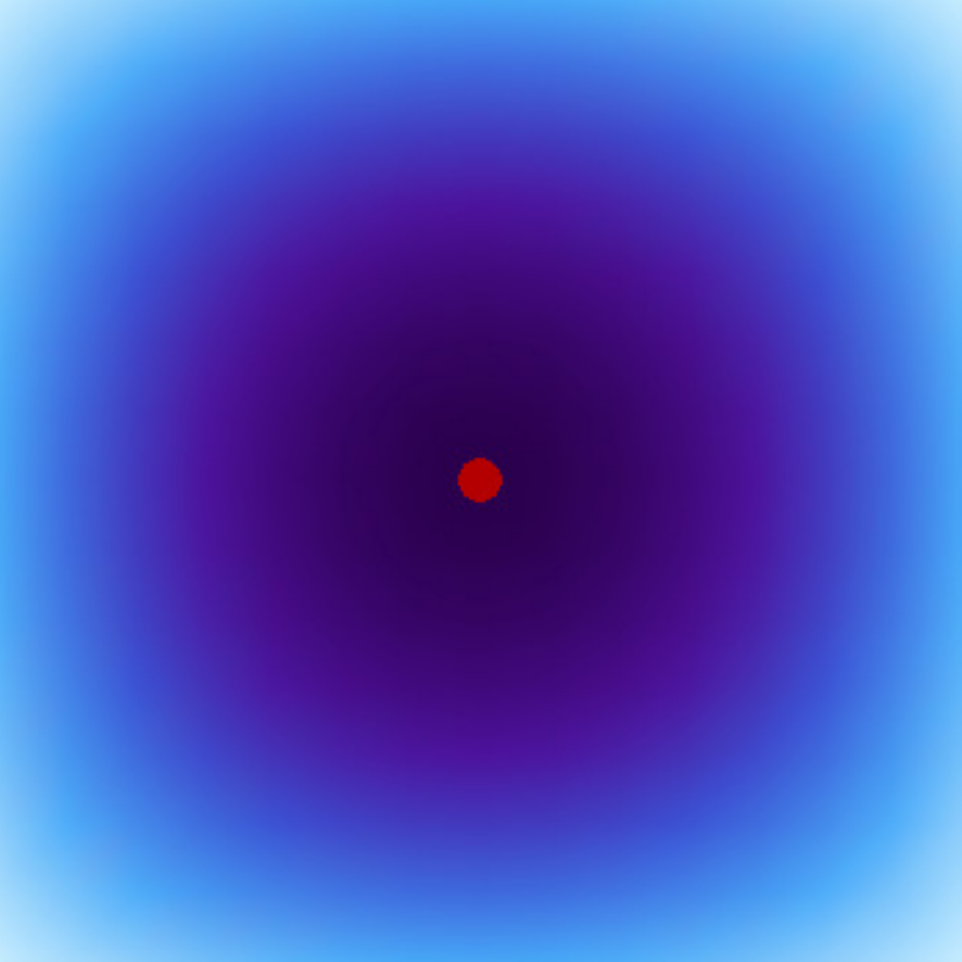}\caption{\small{The clumping of positively charged
fluxes near a negatively charged anti-brane. The blue region corresponds to the flux density and the darker the blue the higher the density.  The red dot represents the anti-brane. }}%
\label{Figure1}%
\end{figure}

In this paper we want to highlight a further problem, already pointed out in \cite{Blaback:2012nf}, that comes to live when one takes into account the back-reaction of the anti-branes. It was first discussed in \cite{DeWolfe:2004qx} that the back-reaction causes a denser 3-form flux near the anti-D3 branes. This is simple to understand. Just like a cloud of 3-branes, a cloud of 3-form fluxes feels a gravitational attraction towards the brane and when the fluxes have the opposite orientation to the 3-brane charge density, the RR forces do not cancel that attraction but instead also pull the fluxes towards the anti-branes. This is depicted pictorially in figure \ref{Figure1} (taken from \cite{Blaback:2012nf}).

The effect of this flux clumping can easily be predicted intuitively. The larger the flux pile-up the higher the probability for the flux to annihilate with the anti-brane. This is very similar to the (thought) experiment in which a positron is dropped into a grid or gas of electrons \cite{Blaback:2012nf}. The higher the electron density the more likely the wave-functions of the positron and electron will overlap to cause annihilation. The back-reaction of the positron only makes things worse by locally creating an increased density of electrons. It is the aim of this paper to provide the equations that back-up this intuition, thereby elaborating on the results in \cite{Blaback:2012nf}. 

The flux clumping effect therefore tightens any bound on stability derived in the probe limit. The essential question  is then whether a vacuum exists at all. For that one needs to compute the flux-clumping and there is by now quite some literature\footnote{\cite{Bena:2009xk, McGuirk:2009xx, Bena:2010gs, Bena:2011hz, Bena:2011wh, Bena:2012bk, Blaback:2011nz, Blaback:2014tfa, Blaback:2011pn, Blaback:2013hqa, Cottrell:2013asa, Dymarsky:2013tna, Giecold:2011gw, Giecold:2013pza, Massai:2011vi,Massai:2012jn}. } on the supergravity solutions that describe back-reacting anti-branes in flux throats, which was initiated by \cite{Bena:2009xk} and \cite{McGuirk:2009xx}. To cut a long story short, it is now well-established that the supergravity solutions have a \emph{diverging} flux density near the anti-brane that in some cases (like the anti-D3 brane in KS) is still integrable.  This can be shown to occur in any background, compact or not \cite{Blaback:2014tfa, Gautason:2013zw}.

At first sight this singular flux is then a killer, but as with all singularities it requires some care to understand them. Roughly speaking singularities can mean two things: 1) There is some short-distance physics that we forgot to take into account that resolves it or 2) the background is genuinely sick for some reason. Of the latter situation we have many examples, such as the negative-mass Schwarzschild solution for instance. It is often also easy to understand in case 2) why the solution is necessarily `sick'. Say one takes a manifestly unstable situation, such as two oppositely charged black holes at a finite distance, and one enforces a static Ansatz to solve the equations of motion. Of course the equations will spit out a solution if you enforce one, but it will contain singularities or closed timelike curves or something else that unveils its unphysical nature. There are some reasons to believe we are in situation 2) with the anti-branes. The `stringy effects' that could cure the anti-D3 singularity can be guessed from the Polchinski-Strassler model \cite{Polchinski:2000uf} as pointed out in \cite{Dymarsky:2011pm}: the singularity should be cloaked by a polarisation proces transverse to the NS5 polarisation. The resulting solution should be a complicated-looking $(p,q)$ 5-brane web. However, all gathered evidence is against that \cite{Bena:2012tx, Bena:2012vz} and is strengthened by a solid criterium for admissible singularities in holographic backgrounds \cite{Gubser:2000nd}, which the anti-brane solutions can be shown not to fulfill \cite{Bena:2012ek, Bena:2013hr, Blaback:2014tfa}. 
 
If true this all indicates that the anti-brane decays classically and then there is no static Ansatz. This also immediately smoothens out the flux divergence as explained in \cite{Blaback:2012nf}.

\section{Branes, throats, and flux clumping}\label{sec:Ansatze}
We briefly review some properties of the supergravity solutions that describe (anti)-branes inside throats. To get a handle on the solutions one necessarily has to make simplifications. At a first level of approximation one ignores the compact manifold in which the throat geometry should be glued into. As long as one studies questions relating to the local description this should be fine. At a second level, simplified throat geometries are studied, such as the Klebanov-Strassler (KS) throat \cite{Klebanov:2000hb}.  In the latter example one can find explicit, numerical anti-brane solutions if the anti-branes are smeared over the $S^3$-tip of the KS throat \cite{Bena:2009xk}.

The metric for a `$p$-brane' throat geometry takes the following form (in Einstein frame)
\be
\mathrm{d}s^2_{10}=e^{2A(\rho)}\mathrm{d}s^2_{p+1}+e^{2B(\rho)} \mathrm{d}s^2_{9-p}\,.
\label{eq:DpAnsatz}
\ee
The worldvolume metric is Minkowski $\d s_{p+1}^2 =\eta_{ij}\d x^{i}\d x^{j}$ and the transversal space is a cone over a base space, $\Sigma_{8-p}$, with  metric
\be\label{metric0}
\mathrm{d}s^2_{9-p}=\mathrm{d}\rho^2+e^{2C(\rho)}\left[ g_{ij}^{\Sigma}\mathrm{d}\psi^i\mathrm{d}\psi^j\right],\\
\ee
where $\rho$ is the radial coordinate and $g^\Sigma_{ij}$ is Einstein;
$R^{\Sigma}_{ij} = (7-p)g^{\Sigma}_{ij}$.

An essential ingredient in throat geometries are the fluxes that support them, and for a $p$-brane throat geometry we have magnetic $F_{8-p}$-flux\footnote{We could equally well use the electric fieldstrength $F_{p+2}$.} 
\begin{equation}
F_{8-p} = \star_{9-p}\d\alpha\,,
\end{equation}
with $\alpha$ a $\rho$-dependent function. This flux can be interpreted as being sourced by actual $p$-branes at the tip of the throat (but this does not need to be the case). Secondly, we have a combination of 3-form flux $H=\d B_2$ and $F_{6-p}$-flux that describe the fractional D($p+2$) branes that have been dissolved into pure fluxes. Hence, these branes carry $p$-brane charge as well (since fractional ($p+2$)-branes do) as can be seen from the Bianchi identity
\be
\d F_{8-p} = \delta_{9-p} + H\wedge F_{6-p}\,. 
\ee
The delta-function describes the explicit $p$-brane sources, if any, and the combination $H\wedge F_{6-p}$ clearly acts as a smooth magnetic source for $F_{8-p}$. In BPS throats the charges induced by the delta function and $H\wedge F_{6-p}$ are of the same sign, meaning that the $(9-p)$-forms $\delta$ and $H\wedge F_{6-p}$ have the same orientation. In the latter case the fluxes furthermore obey a Hodge duality relation
\cite{Blaback:2010sj}
\be \label{dualitycondition}
\star_{9-p}H_3 =  e^{\frac{p+1}{4}\phi}F_{6-p}\,,
\ee
where the Hodge star $\star_{9-p}$ is taken with respect to the whole metric transversal to the worldvolume (so including warp factors).
When $p=3$, this is known as the ISD condition. In such backgrounds a D$p$ probe feels no force and when inserted into the throat its backreaction will not alter the profile of the background fluxes $F_{6-p}$ and $H_3$.

In this paper we care about the opposite situation in which the $\delta$-source orientation is opposite to that of the fluxes. The $H\wedge F_{6-p}$-term dynamically changes over the throat and its orientation can flip. Hence, to define more precisely what it means to `put an anti-brane down the throat',  we demand that the orientation of the term $H\wedge F_{6-p}$ far away from the tip (the UV) approaches (\ref{dualitycondition}) and that it is opposite to the delta function at the tip. As a consequence (\ref{dualitycondition}) does not hold anymore inside the throat. Instead it has been shown, \emph{regardless of any smearing or symmetry argument}, that the $H$-flux will diverge near the tip as follows \cite{Blaback:2014tfa, Gautason:2013zw}:
\begin{equation}\label{H^2}
e^{-\phi}H_3^2\sim e^{-2A}\,.
\end{equation}
This is a divergent,  but integrable, scalar, since near the brane source $e^{2A}$ approaches zero. As mentioned earlier, this singularity has been interpreted \cite{Blaback:2011nz, Blaback:2010sj} to be a consequence of `flux clumping' \cite{DeWolfe:2004qx}: the gravitational attraction between branes and fluxes is not counterbalanced anymore with the RR repulsion since the RR forces are also attractive for anti-brane sources. This interpretation remains somewhat under appreciated in the literature. We stress that this interpretation is strengthened by the fact that the 3-form singularity blows up with such an orientation in the form $H\wedge F_{6-p}$, that it corresponds to a pile-up of D3 brane charges dissolved in flux. If the singularity would have a different sign of $H\wedge F_{6-p}$, this interpretation would be incorrect and all of its consequences discussed below, would not hold.

The $H$-flux always has a part along $\star_{9-p}F_{6-p}$ and it is exactly this part that necessarily becomes singular (the other directions can also blow up) since that part contributes positively to the charge density in $H\wedge F_{6-p}$. The coefficient of $H$ along that direction is then defined to be the \emph{flux clumping} parameter $\lambda$ as follows
\be\label{lambdadefinition}
\star_{9-p}H_3 = \lambda e^{\frac{p+1}{4}\phi} F_{6-p}\,,
\ee
For certain simple throat geometries (e.g.~Klebanov-Tseytlin ) the only legs of $H$ are along $\star F_{6-p}$ and then the flux clumping parameter can describe the full clumping.

\section{Enhanced brane-flux annihilation}\label{sec:enhanced}

The physical picture of brane-flux annihilation suggests that the pile-up of flux near the anti-brane enhances the annihilation process. In this section we roughly quantify this by tracing back how the flux-clumping parameter $\lambda$ enters the stability conditions and decay rate\footnote{The philosophy behind this rough description of the true flux clumping is best argued with an analogy: to understand whether a candle goes out when put close to the waterline at the beach, one does not need to compute the complicated shape of the water waves, one simply needs to know whether a wave is coming.}. We present three different viewpoints, as done in the original KPV paper \cite{Kachru:2002gs}: 1) we consider the Abelian NS5 worldvolume action, 2) the non-Abelian wordvolume theory of $p$ anti-D3 branes  and 3) the nucleation of bubbles of supersymmetric vacuum in 4d space-time. The effect of flux-clumping from the NS5 viewpoint was already discussed in \cite{Blaback:2012nf} and we recall those results here for completeness. 

The approach we follow consists in improving on the probe computations of \cite{Kachru:2002gs}. In the probe limit the supergravity solution is pure Klebanov-Strassler. In what follows we need the near-tip geometry which is given by (in the notation of \cite{Kachru:2002gs})
\begin{align}
& \d s^2 = a_0^2 \eta_{\mu\nu}\d x^{\mu}\d x^{\nu} + g_s M b_0^2 (\tfrac{1}{2}\d r^2 + \d\Omega_3^2 + r^2\d\Omega_2^2)\,,\nonumber\\
& F_3 = M\Omega_3 \,, \qquad H_3 = g_s\star_6 F_3\,, \nonumber\\
& C_4 = 0 \,,\label{probegeometry}
\end{align}
where
\begin{equation}
a_0^2 =\frac{\epsilon^{4/3}}{g_s M}\,,\qquad  b_0^2\simeq 0.93266 \,,
\end{equation}
and the normalisation of the volume form $\Omega_3$ is such that
\begin{equation}
\int_{S^3}F_3 = 4\pi^2 M \,.
\end{equation}

We improve on the probe computation by considering the effect of the flux-clumping parameter $\lambda$ (\ref{lambdadefinition}). So we will trace back how $\lambda$ enters the probe computation, since it is assumed to be equal to one in \cite{Kachru:2002gs}. In subsection \ref{consistency} we comment on the consistency of this approach.

\begin{figure}[h] \centering
\includegraphics[width=.6\textwidth]{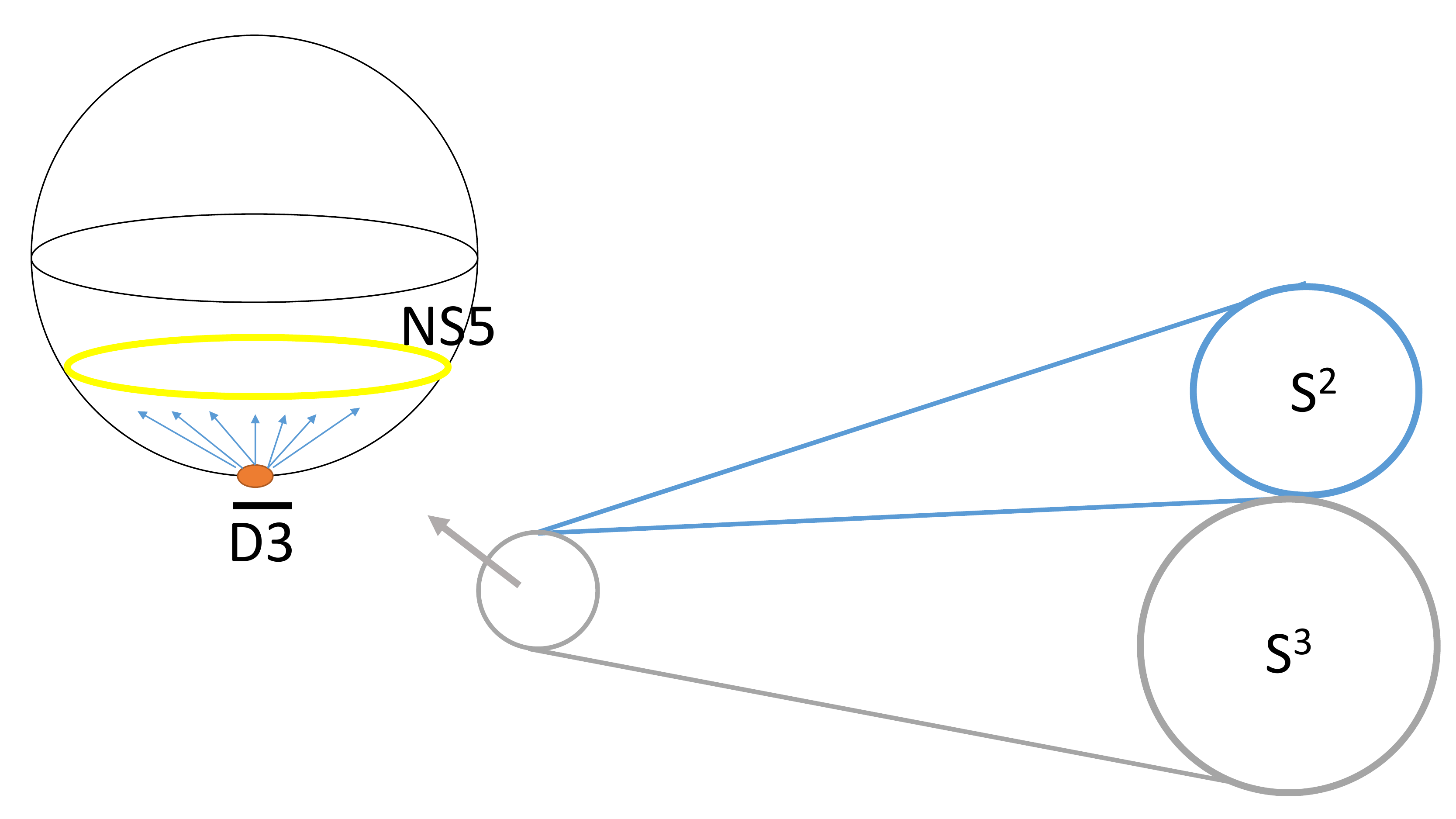}\caption{\emph{The polarisation of the anti-D3 brane into a spherical NS5-brane wrapping a contractible $S^2$ inside the $S^3$ at the tip.} }%
\label{Polarisation}%
\end{figure}

Brane-flux decay proceeds through the Myers effect \cite{Myers:1999ps} as was found in \cite{Kachru:2002gs}. Due to the 3-form fluxes the anti-D3 brane at the tip will polarise into a spherical NS5-brane that wraps a contractible $S^2$ in the finite-size $S^3$ at the tip, see figure \ref{Polarisation}.
Depending on the position of this $S^2$ the spherical NS5-brane carries $-p$ units of D3 charge (ie $p$ anti-D3 branes) or $M-p$ units of D3 branes (since $p<<M$ we have that $M-p>0$): $-p$ D3 charges are induced when the $S^2$ pinches of at the South Pole and $M-p$ units are induced when it pinches of at the North Pole. This does not violate charge conservation since at the same time the transition from South to North Pole occurs, the 3-form flux associated to $H_3$ drops one unit such that the total D3 charge induced by the 3-form fluxes makes up the difference. Hence the proper way to interpret the transition from South to North Pole is the materialisation of $M$ D3 branes out of the 3-form fluxes. Subsequently those $M$ D3 branes find the $p$ anti-D3 branes after which $M-p$ D3 branes are left over. This end stage is the (mesonic branch of) SUSY vacuum, ie, the KS geometry with mobile D3 branes. The process that makes one unit of $H_3$-flux drop is bubble nucleation a la Brown-Teitelboim \cite{Brown:1987dd}, which is nothing but a stringy generalisation of the Schwinger effect. The bubble is then a different spherical NS5-brane that has one co-dimension inside the external space and it wraps the entire $S^3$. However the two NS5-branes, the one wrapping an $S^2$ inside $S^3$ and the one wrapping the whole $S^3$ can be thought of as the two ways to view the same process of brane-flux decay. Effectively the NS5-brane that wraps the $S^2$ can be seen as a stringy resolution of the thin wall of the bubble. The figure in subsection \ref{sec:bubble} illustrates this and more comments can be found there.

\subsection{The NS5 point of view}

We now compute the potential energy of a spherical NS5-brane wrapping the contractible $S^2$ inside the $S^3$ at the tip of the KS throat. This potential tells us about the dynamics of the NS5 motion. A meta-stable state would imply that the NS5-brane does not get pushed immediately to the North Pole but that there is a classical barrier for this. The force on the NS5-brane towards the North Pole is due to the 3-form flux. In particular it is proportional to the integral of $B_6$. The flux pile up implies that this force grows with $\lambda$ as we now explicitly check.

The NS5-brane has a worldvolume flux of the Abelian Born-infeld vector that induces the 3-brane charge as can be seen via the coupling to $C_4$ in the Chern-Simons term
\be\label{CSACTION}
\mu_5 \int B_6 + 2\pi\mathcal{F}_2\wedge C_4\,.
\ee
The first term is standard and describes how the NS5-brane is charged electrically w.r.t~$B_6$. This contribution is sensitive to the flux clumping, since, with the help of (\ref{lambdadefinition})\footnote{We have inserted a minus sign wrt to equation (\ref{dualitycondition}) in order to follow the sign conventions of \cite{Kachru:2002gs}. }, we find
\begin{equation} \label{B6}
\d B_{6}\equiv\frac{1}{g_{s}^{2}}\star_{10}H=-\frac{\lambda}{g_{s}}V_{4}\wedge
F_{3}\,.
\end{equation}
Here $g_{s}$ is the string coupling and $V_{4}$ is the red-shifted volume-form, along the four non-compact dimensions. 
The second term in the Chern-Simons action (\ref{CSACTION}) contains $\mathcal{F}_2$ which is defined as
\be
2\pi\mathcal{F}_2 = 2\pi F_2 -C_2\,,
\ee
where $F_2$ the field strength of the BI vector and $C_2$ the RR gauge potential.
Since $F_{3}$ is
topologically ``protected'' to have $M$ units of flux around the
$S^{3}$, we preserve the expression for $C_{2}$ from the KS
solution. This gives
 \cite{Kachru:2002gs}
\begin{equation} \label{C2 integral}
\int_{S^{2}}C_{2} = 4 \pi M (\psi-\tfrac{1}{2}\sin(2\psi))\,,
\end{equation}
where $\psi$ is the third Euler angle of the $S^{3}$ that measures
the sizes of the various $S^{2}$'s within $S^{3}$. Since the
NS$5$-brane wraps these $S^{2}$'s and moves along on the $S^{3}$,
$\psi$ is used to keep track of the position of the NS$5$-brane. 
The fact that the NS$5$-brane induces, initially, $p$
anti-D$3$ charge is due to the monopole charge of the worldvolume
flux $F_{2}$ of the NS5-brane, through the $S^{2}$
\begin{equation}\label{F2integral}
2\pi\int_{S^{2}} F_{2} = 4\pi^{2} p\,.
\end{equation}
Using equations (\ref{C2 integral}, \ref{F2integral}) one easily verifies that at the North Pole, where
$\psi=\pi$, the NS5  induces $M-p$ D$3$ charges instead of $p$ anti-D$3$
charges at the South Pole.
 
\begin{figure}[h] \centering
\includegraphics[width=.6\textwidth]{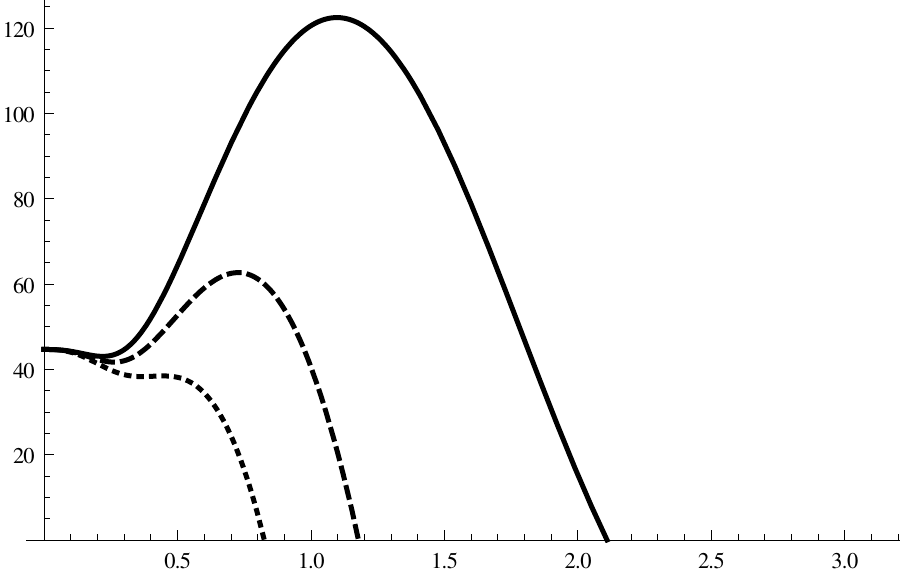}\caption{\emph{The
effective potential for the motion of the NS$5$ brane, plotted for
different
values of $\lambda$. The vertical axis denotes the value for the potential and the horizontal axis the Euler angle $\psi$. This picture taken from \cite{Blaback:2012nf}}. }%
\label{Figure2}%
\end{figure}

If one regards the dynamics of the NS$5$-brane motion as described by the effective one-dimensional action for a time-dependent $\psi$
\cite{Kachru:2002gs} one can deduce the potential energy for the NS5 position from summing the DBI action
\be \label{DBI}
\mu_5 g_s^{-2} \int_6 \sqrt{\text{det}(G_{//})}\sqrt{\text{det}(G_{\perp} + 2\pi g_s\mathcal{F}_2)}\,,
\ee
with the CS term (\ref{CSACTION}) to find:
\begin{equation}
\label{potential} V_{eff} (\psi) \sim \frac{1}{\pi} \sqrt{b_{0}%
^{4}\sin^{4}\psi+ \bigl(\pi\frac{p}{M}-\psi+\tfrac{1}{2}\sin(2\psi)
\bigr)^{2}} - \frac{\lambda}{2\pi}(2\psi-\sin(2\psi))\,,
\end{equation}
The effect of $\lambda$ in the
potential is made clear in the three plots in figure \ref{Figure2}, taken from \cite{Blaback:2012nf}.
The plot for $\lambda=1$ (full line) shows the meta-stable
vacuum of \cite{Kachru:2002gs}, with $p/M$ chosen as $3\%$. The dashed line
corresponds to $\lambda\approx1.3$ , in which the vacuum is almost gone and the third plot, with dotted line, for which
$\lambda\approx1.7$ shows no meta-stable vacuum any more. Lower values of $p/M$ increase the height of the barrier and to ensure that the stable minima will disappear for very small values of
$p/M$, $\lambda$ needs to be of the order $(p/M)^{-1/2}$ \cite{Blaback:2012nf}, as reviewed below. 

If we focus on small values of $p/M$, where the critical value $\lambda_c$ for the flux clumping is large, it is possible to obtain an analytic expression for $\lambda_c$ as was done in \cite{Blaback:2012nf}. In this limit we have $\psi \sim \epsilon, \lambda \sim 1/\epsilon$ and $p/M \sim \epsilon^2$, with $\epsilon$ small. We find
\begin{equation}
\label{potential} V_{eff} (\psi) \sim \frac{1}{\pi} \sqrt{b_{0}^{4}\psi^4+ \bigl(\pi\frac{p}{M}\bigr)^{2}} - \frac{2\lambda}{3\pi}\psi^3\,,
\end{equation}
from which it is easy to derive the critical value at which the meta stable minimum disappears \footnote{This corrects a harmless typo in \cite{Blaback:2012nf}.}:
\be \label{lambdacritical1}
\lambda_c \simeq \frac{b_0^3}{\sqrt{2\pi}}\Bigl( \frac{p}{M}\Bigr) ^{-1/2}\,.
\ee
We refer to the appendix for some more details on the calculation. Sometimes it useful to  consider $\lambda/b_0^3$, which is a density on the $S^3$ on which the NS5 can wrap, rather than $\lambda$ itself. The critical value for this density is of order $\Bigl( \frac{p}{M}\Bigr) ^{-1/2}$.

One can also expand the potential for small $\psi$ to find
\be\label{potentialexpanded}
V_{eff} \sim  \frac{p}{M} -\frac{2}{3M} (1+\lambda)\psi^3 + \frac{1}{2}b_0^4\frac{M}{\pi^2 p}\psi^4\,.
\ee
Written in this way, one might get the impression that there always is a minimum, but one easily checks that the expansion breaks down for values of $\psi$ near this would be minimum when $\lambda$ is of the order of the critical value in (\ref{lambdacritical1}).  We will return to this expansion later on when comparing with the results of the non-Abelian D3 action and bubble nucleation.

\subsection{The anti-D3 point of view}
The non-Abelian worldvolume action for $p$ anti-D3 branes \emph{in the S-dual frame} is given by
\be
S = \frac{\mu_3}{g_s} \int\text{Tr}\sqrt{\text{det}(G_{||} + 2\pi g_s F_2) \text{det}(Q)} -2\pi\mu_3 \int \text{Tr} i_{\phi}i_{\Phi}B_6\,,
\ee
where
\be
Q_{i}^{j} =\delta_{i}^j +\frac{2\pi i}{g_s}[\Phi^i, \Phi^k](G_{kj} + g_s C_{kj} )\,.
\ee
In the above $G_{||}$ denotes the worldvolume metric at the apex of the deformed conifold, which we can locally approximate to be flat $G_{||} \sim \delta$. The $\Phi^i$ are the non-commutative scalars whose eigenvalues denote the classical position of the brane inside the conifold. In the below we use the convention that repeated scalar indices $i$ are summed over, regardless of their height.
 
 The 2-form RR gauge potential $C$ can be expanded at first-order in the $\Phi$ as
\be
C_{ij} \simeq F_{ijk}\Phi^k=f\epsilon_{ijk}\Phi^k \,,\quad\text{with}\quad f \simeq \frac{2}{b_0^3\sqrt{g_s^3M}}\,. 
\ee
For $B_6$ we apply the above Ansatz (\ref{B6}). Putting all of this information together one finally finds the following potential for the scalar $\Phi$ up to quartic order:
\be
g_s V(\Phi) = p - 2 i \pi^2 f \frac{(1+\lambda)}{3}\epsilon_{kjl}\text{Tr}\Bigl([\Phi^k,\Phi^j]\Phi^l\Bigr)-\frac{\pi^2}{g_s^2}\text{Tr}\Bigl([\Phi^i, \Phi^j][\Phi_i, \Phi_j]\Bigr)\,.
\ee  
We ignore the constant term in this potential, although it plays an interesting role as ``uplifting energy''.
The KPV result is obtained by simply putting $\lambda=1$. As before the strategy is to infer what the effect of flux clumping, ie large $\lambda$, amounts to. For that we compute the stationary point of this potential, from the extremality condition
\be
[[\Phi^i, \Phi^j], \Phi^j]  - i g_s^2 f\frac{(1+\lambda)}{2}\epsilon_{ijk}[\Phi^j, \Phi^k]=0\,.
\ee
This can be solved by the following $SU(2)$ algebra
\be
[\Phi^i,\Phi^j] = i g_s^2 f\frac{(1+\lambda)}{2}\epsilon_{ijk}\Phi^k\,,
\ee
which describes a fuzzy 2-sphere with effective radius
\be \label{radius}
R^2 =\frac{4\pi^2}{p}\text{Tr}[(\Phi^i)^2] \simeq \pi^2g_s^4 f^2 \bigl(\frac{1+\lambda}{2}\bigr)^2 \frac{c_2}{p} = \frac{4\pi^2}{b_0^8M^2}\bigl(\frac{1+\lambda}{2}\bigr)^2 \frac{c_2}{p} R_0^2\,,
\ee
with $R_0$ the radius of the $S^3$: $R_0  = b_0 \sqrt{g_s M}$ and $c_2$ is the value of the quadratic Casimir
\be
c_2 = \sum_{i=1}^3\text{Tr}[J_i]^2\,.
\ee
One can verify that the lowest energy configurations are such that the $J_i$ correspond to the $p$-dimensional irreducible representations of $SU(2)$, hence $c_2 = p(p^2-1)$.

Again we can identify the same critical value for $\lambda_c$ as we have calculated before. At this value the potential $V$ vanishes and becomes negative for larger $\lambda$, as can be seen from 
\be
V \simeq \frac{\mu_3}{g_s}p\Bigl(2 - \frac{8\pi^2 (p^2-1)\lambda^4}{3b_0^{12}M^2} \Bigr)\,.
\ee
This can be understood as the same kind of breakdown of the expansion up to quartic terms, as we saw around equation (\ref{potentialexpanded}) and it signals the disappearance of the meta-stable minimum.

\subsection{Bubble nucleation} \label{sec:bubble}

We have argued that there will be a build up of large, potentially diverging, flux on top of the anti-branes spanning space-time. This is a simple consequence of flux being attracted towards the anti-branes when supersymmetry is broken, and the gravitational and RR-forces no longer cancel. To be more precise, it is the $H_3$ part of the flux that will diverge, and we have also argued that the presence of this large field strength on top of the anti-branes will lead to a perturbative instability, and trigger the annihilation of the anti-branes. We will now consider this process using the analogy with the Schwinger effect. 

The usual Schwinger effect involves pair creation of charged particles in the presence of a strong electric field. Two massive particles of opposite charge, can in principle be created without a cost in energy if they start out so close to each other that the negative potential energy due to their attraction balances their masses. They can, however, never escape to infinity unless there is an external electric field. If the two particles are moved away from each other, the total energy will increase until the external field overcomes their mutual attraction, after which the total energy will again decrease. Hence, there is a barrier that needs to be overcome. The probability of tunneling through this barrier is interpreted as the probability of spontaneous pair creation.

The higher dimensional analog of the pair creation described above is the creation of neutral, expanding shells containing new vacua. The probability of a bubble to form is given by
\be
P \sim e^{-B/\hbar},
\ee
where
\be
B=S_E[{\rm instanton]} - S_E[{\rm background]},
\ee
is the difference in the Euclidean action between a background with or without the bubble. 

We will focus on the limit where the probability to tunnel becomes large, and the size of the bubbles small compared to the de Sitter radius, such that effects of gravity can be ignored. In the limit of thin walls, the Euclidean action for creating a bubble of radius $R$ is then given by
\be
S_E=-\rho V_{d+1}+\sigma A_{d},
\ee
while the energy of the created bubble is
\be
E=-\rho V_{d}+\sigma A_{d-1}.
\ee
Here, $\sigma$ is the tension of the bubble wall, and $\rho$ the difference in energy between the inside and the outside of the bubble, while $V_d =\frac{2\pi^{d/2}}{d\Gamma [d/2]} R^d$ and $A_{d-1} =\frac{2\pi^{d/2}}{\Gamma [d/2]} R^{d-1}$
are the inside volume and area of $S^{d-1}$ respectively. Minimizing the action tells us that the radius of the created bubble is given by
\be
R=\frac{d\sigma}{\rho},
\ee
which also is the radius for which the energy of the created bubble vanishes, in line with energy conservation. For $d=3$ this gives the tunneling amplitude
\be
P \sim e^{-\frac{27\pi^2 \sigma^4}{2\hbar \rho^3}}.
\ee
In our case it is the $H_3$-field that couples to NS5-branes and will induce the spontaneous creation of shells of NS5 inside of the anti-D3's. We will now investigate in detail what happens when these bubbles form.

\begin{figure} \center
\includegraphics[height=70mm]{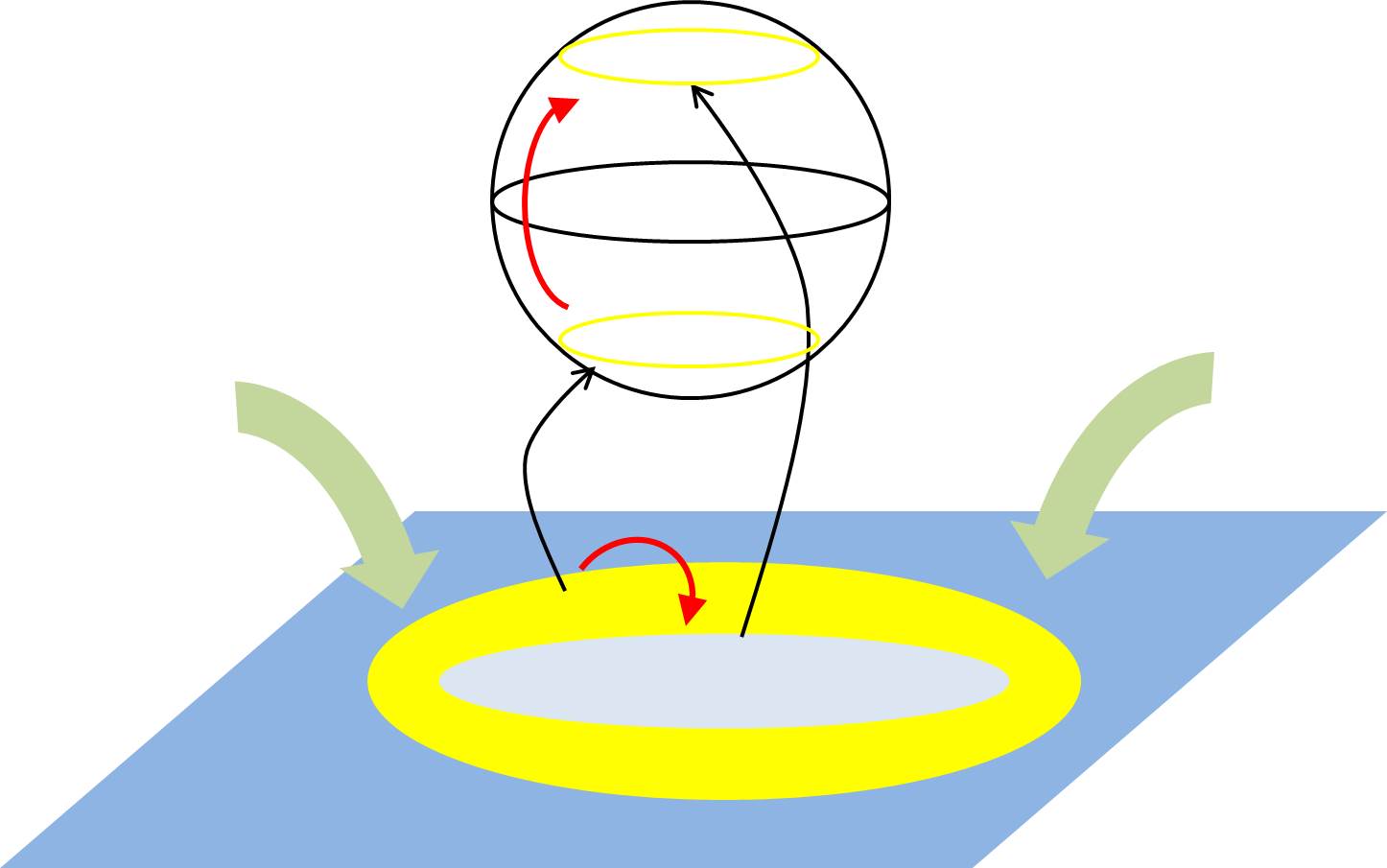} 
\caption{\emph{The three dimensions of the NS5-brane transverse to the bubble wall correspond to the thickness of the wall and an internal $S^2$. The $S^2$ moves across the $S^3$ when you go through the bubble wall.}. }%
\label{bubble}%
\end{figure}

An anti-D3 can end on the NS5, but there will be three extra dimensions left over. That is, a bubble in the anti-D3 will have two directions parallel to the bubble wall, and three transverse directions. Two of these transverse directions will wrap an $S^2$ inside of the $S^3$, while the third direction will correspond to the thickness of the bubble wall in space-time. As you move from the outside edge of the bubble wall across to the inner edge, the internal $S^2$ will, as it moves in the coordinate $\psi$, expand from one of the poles on the $S^3$, go over the equator, and then collapse towards the other pole. Hence, in the infinitely thin wall limit this spherical NS5-brane fills or wraps the whole $S^3$ at once and therefore it might naively look as if there are two different NS5-branes, the Brown-Teitelboim one and the puffed anti-D3, but they are really the same. In the opposite limit, the infinitely thick wall limit, the NS5-brane fills the whole 4d space and wraps the $S^2$ inside $S^3$. We have illustrated this point of view in figure \ref{bubble}. 

In this process  the integrated charged fluid density $ \oint H \wedge F_3 \sim KM$ will change to $KM-M$ within the bubble, since the $H$-flux within the bubble is one unit smaller due to the presence of the NS5, which sources the $H$. What happens is that part of the fluid is converted into $M$ D3-branes. These D3-branes are then free to annihilate against the anti-D3 branes within the bubble. In this way bubble nucleation make the anti-branes go away in a process completely analogous to the Schwinger effect.

If you remain at a fixed point in space, waiting for the expanding bubble to come, the transition will proceed trough a time-dependent process as the wall moves past, where the anti-D3 are puffed up to an $S^2$ that moves down (possibly through tunnelling) in the potential of figure 3.

As we have seen this process can proceed perturbatively if the field strength is big enough. The reason is that the internal structure of the wall of the bubble is resolved within SUGRA, as we effectively move beyond the crudest thin wall approximation. As $\lambda$ is increased the effective tension is reduced, and at the critical value $\lambda _c$ there is no barrier left, and no bubble wall with tension. Even though the tension of the five dimensional NS5-brane never vanishes, the thick wall of the two dimensional bubble has an effective tension given by a combination of the NS5-tension and the effect of the strong $H$-field. This should be contrasted with the case of electron-positron pairs where the mass of the particles stay constant, and there always is a tiny little barrier left to tunnel regardless of how strong the field is. 

In the appendix we calculate the effective tension of the bubble as we approach the critical value for $\lambda$. The result is
\be
\sigma = \frac{3}{5} 2^{19/8} \sqrt{\pi g_s M} (\frac{p}{M})^{3/2} (1-\frac{\lambda}{\lambda _c})^{5/4} \,.
\ee
This value is only valid when $\lambda$ is close to $\lambda_c$. With this we can explicitly see how the suppression due to tunnelling disappears as the critical value of $\lambda$ is approached. In case back-reaction is ignored the same computation would instead give exactly the tension of a probe NS5-brane wrapping the whole $S^3$ \cite{Kachru:2002gs}.

\subsection{Consistency of the approach} \label{consistency}

At first sight one might have worries about the approach we have followed. We therefore address two questions we believe could prompt the worried reader: 
1) why did we only consider flux clumping as a back-reaction effect? 
And, 2) how can we make any conclusions about anti-brane solutions 
since $\lambda$ does become infinite in that case? Question 1) also 
relates to the objection that the followed approach implies
``filling in the back-reaction of the probe into the probe action", which 
is known to be inconsistent (e.g.~the self energy of an electric 
source needs to be subtracted in the coupling $j_{\mu} A^{\mu}$).

We start by addressing the first question. A probe action should 
indeed only be used for a probe source. A consistent method for 
taking into account back-reaction into the probe action works as follows: one forgets about the quantisation of charge and tension and considers the $p$ anti-D3 branes to consist of $p-\epsilon$ back-reacting anti-D3 branes and $\epsilon$ probe branes 
with $\epsilon$ small
\be
 p\, \overline{D3}\quad =\quad (p-\epsilon)\underbrace{\overline{D3}}_\text{back-reacting}\quad + \quad \epsilon\,\underbrace{\overline{D3}}_\text{probe} \,.
\ee
One then investigates whether the $\epsilon$ probe branes are meta-stable in the background of the $p-\epsilon$ back-reacting branes. If the latter are unstable one takes it as a sign that the whole stack is unstable (recall that the higher $p$ the less stable anti-D3 branes are). Hence one can fill in the back-reacted solution into the probe action, but one has to keep in mind what one is doing. 

\begin{figure} \center
\includegraphics[height=70mm]{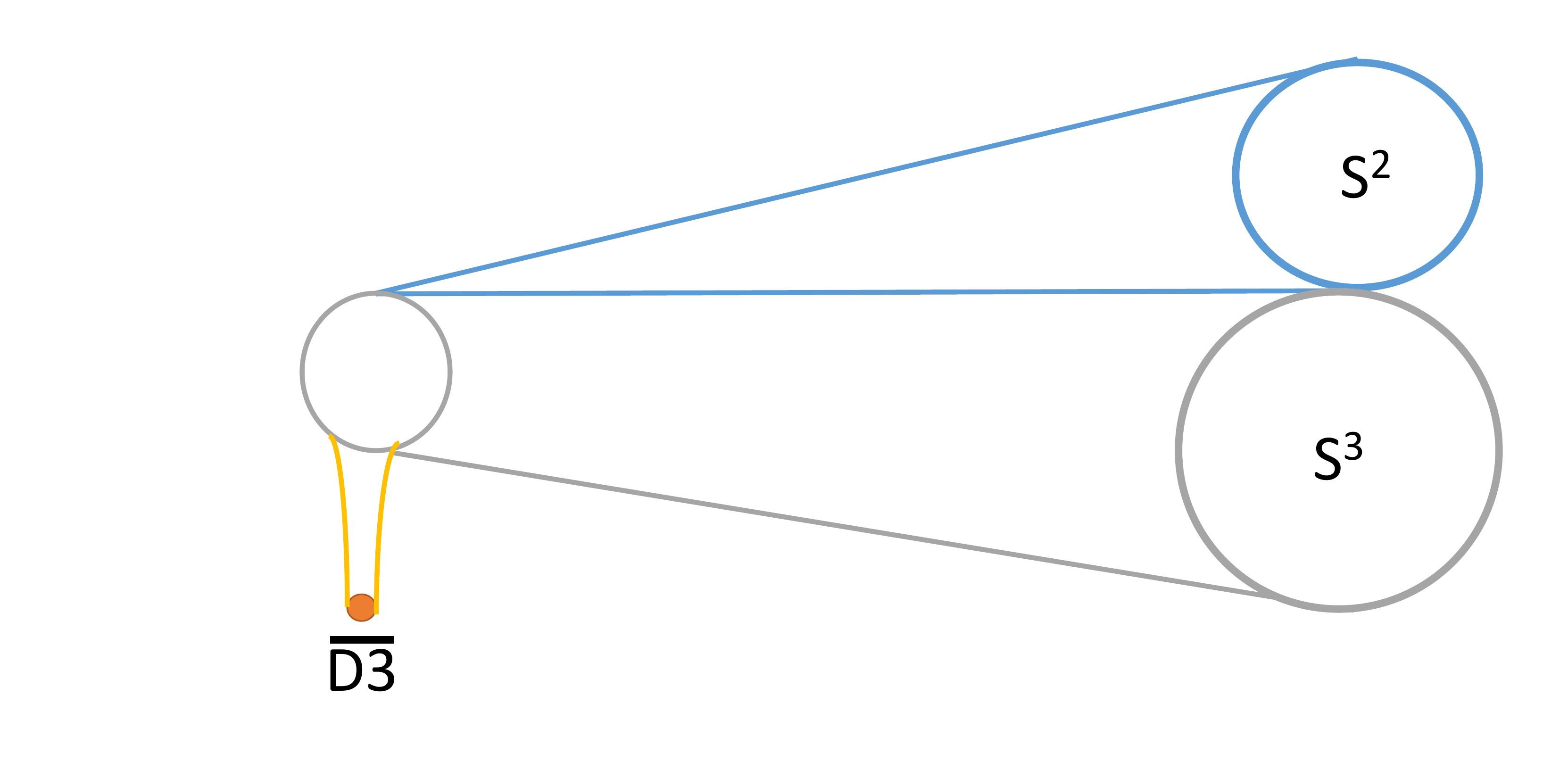} 
\caption{\emph{At the tip of the KS throat a new AdS throat opens up around the anti-D3 branes}. }%
\label{AdSthroat}%
\end{figure}

The back-reaction on the geometry is expected to be such that at the tip of the KS throat we generate a new, much thinner throat, which is locally $AdS_5\times S^5$ as illustrated in figure \ref{AdSthroat}.
The throat geometry causes the space to be locally stretched and one might think that this could cure the enhanced brane-flux decay. After all, the $S^3$ becomes bigger and it will be more difficult to push the NS5-brane over the equator. Put differently, the local blow up of the fluxes goes together with a local stretching of the space and the two effects might cancel out.

There is a simple qualitative picture of what happens as we move down into the local AdS throat, which suggests that such a cancellation does not take place. We already know, from \cite{Blaback:2014tfa, Gautason:2013zw}, that the flux clumping is a completely general result, and is not removed through the presence of the local AdS throat. But if we want to discuss the stability of a probe with respect to it being pushed over the equator of the $S^3$ towards the SUSY vacuum, this is not enough and we must also understand what the effect is on the local stretching (warping) of space-time. We did not consider this effect in \cite{Blaback:2012nf}. Qualitatively the geometry (\ref{probegeometry}) felt in the probe regime should be corrected as follows, at leading order,
\begin{equation}
a_0^2 \rightarrow a_0^2 \exp(2A)\,,\qquad b_0^2 \rightarrow b_0^2 \exp(-2A)\,,
\end{equation}
and 
\begin{equation}
 H_3 = \lambda g_s\star_6 F_3\,. \nonumber
\end{equation}
 Consider now for instance the equation for the critical value of $\lambda$ (\ref{lambdacritical1})
\be 
\lambda_c \simeq \frac{b_0^3}{\sqrt{2\pi}}\bigl(\frac{p}{M} \bigr)^{-1/2} \sim \exp(-3A)\,.
\ee
We notice that $\lambda_c$ grows as $\exp(-3A)$, whereas $\lambda$ grows as $\exp(-4A)$ and therefore $\lambda$ is always expected to grow over $\lambda_c$ down the throat. In other words, a growing $\lambda$ beats the stretching effect when it comes to the force felt by the NS5-brane that pushes it over the equator towards the North Pole. 

There is yet another effect that we need to discuss on top of what we did in \cite{Blaback:2012nf}. Due to the localization $\lambda$ will, in general, be a function that depends on the distance from the equator and the positions of the anti-branes or the puffed up NS5. As the branes start to move one can even expect a time lag before the flux catches up.  However, given that our goal is to analyze whether there exist a static solution, it is self consistent to focus on the adiabatic limit where the flux simply follows the position of the branes. In this sense one can still talk about a specific potential depending on $\psi$ and parametrized by the flux clumping parameter as we used above. 

To fully grasp the effects we have discussed one would have to study the AdS throat by itself, where one gets a, local, non-supersymmetric version of the Polchinski--Strassler model \cite{Polchinski:2000uf}. Unfortunately, such an analysis can only be carried out in a regime where $p>>M$, which is the regime opposite to the regime of interest where $p<<M$. Nonetheless, this regime has been studied in the recent work by Bena et al, \cite{Bena:2014jaa}, and the results clearly indicate the would-be vacuum suffers from tachyonic instabilities. The end-point of these tachyonic instabilities cannot be guessed from \cite{Bena:2014jaa} but we suggest that the end-point is the most obvious one: the SUSY KS solution with, possibly, some left over mobile D3 branes. Inside the ``Polchinski-Strassler'' regime of space-time the $S^3$ geometry at the tip is invisible and lives in the far UV. Nevertheless, it is reassuring that the results point in the same direction in this approach. A different way to understand why we are in a regime complementary to the analysis in \cite{Bena:2014jaa}, comes from the simple observation that in our analysis the instability of the probe arises at the same point as the expansion of the polarisation potential to quartic order breaks down. The Polchinski-Strassler analysis of \cite{Bena:2014jaa} relies on such an expansion to be meaningful. It would be very interesting to provide a unified picture for all values of $p$. We conjecture that the underlying process behind the instability in all cases is flux annihilation. 

Now we address the second question concerning the actual divergence of $\lambda$. In any real process the flux density can never diverge. Instead, one has to regard the formally diverging value as a sign that something is not yet understood. One logical possibility is that stringy effects, such as brane polarisation of $(p,q)$ 5-branes, or something else can resolve the singularity. In the discussion we comment in more detail on the possibility of a resolution of the singularity that would save the existence of the KPV meta-stable state. Let us for now mention that the collected evidence is against such a resolution \cite{Bena:2012tx, Bena:2012vz, Bena:2014bxa, Bena:2012ek, Bena:2013hr, Blaback:2014tfa}.  It is the opinion of the authors that the most plausible possibility is that the singularity indicates a perturbative instability \cite{Blaback:2012nf}. The system is ``side of the hill''. In the latter case we must take time-dependence into account. At some time $t=t_{0}$ an anti-D3 brane is dropped into the KS throat. Subsequently the RR and gravitational forces pull the flux towards the anti-brane and $\lambda$ is effectively growing. It grows until it reaches the critical density $\lambda_c$ which sets off the classical brane-flux annihilation that puts a stop to the flux accumulation near the anti-D3 brane. This process occurs until all of the anti-brane charge is annihilated and one is left over with the SUSY vacuum at late times. This sidesteps the discussion of the singularity since there is no singularity at any moment in time for such a time-dependent background. So ``time-dependence'' can be seen as a resolution of the singularity \cite{Blaback:2012nf}.

\section{Discussion}\label{sec:disc}
As explained in the previous section, the classical instability that occurs at $\lambda > \lambda_c$ implies that the singular static  anti-brane backgrounds should really be time-dependent in such a way that $\lambda$ can never grow indefinitely and hence time-dependence resolves the puzzling 3-form singularity at the expense of the loss of the meta-stable KPV state  \cite{Blaback:2012nf}. In what follows we first explain how our results imply that a different resolution of the singularity, that preserves the meta-stable state, is more unlikely. We distinguish between a resolution at length scales that can be described in supergravity and at length scales that are beyond the supergravity approximation. Before we conclude at the very end we comment on some arguments in favor of the KPV state.

\subsection{Can the singularity be resolved within SUGRA?}

It seems sensible that the only possible resolution of the singularity should be one within the supergravity approximation, at large radius.  Imagine the contrary, that somehow $\alpha'$ corrections are able to smoothen out the singularity. If so, the supergravity solution will only get substantially altered at a radius of the order of the string scale. This would then be inconsistent with the above constraints from brane/flux annihilation since the flux density $\lambda$ already grows over its critical density $\lambda_c$ at a radius much larger than the string scale. Once the flux density crosses the critical density, nothing can prevent brane-flux annihilation to occur, as was shown first in \cite{Blaback:2012nf} and elaborated upon in this paper. The intuition behind this is clear. The singular three-form flux can be seen as a local pile of 3-brane charges dissolved in flux. If this `cloud' of D3-branes comes too close to the anti-brane, a direct annihilation occurs. This is why there exists a critical value for the flux density required for stability, and this is why one wants the singular solution to be corrected at large enough scales in order to stay below this critical density.

Brane polarisation \cite{Myers:1999ps, Polchinski:2000uf} is the natural candidate for a resolution at large scales within supergravity, but it has so far not been shown to work \cite{Bena:2012tx, Bena:2012vz, Bena:2014bxa}. This is in line with nogo results for a smooth solution at finite temperature \cite{Bena:2012ek, Bena:2013hr, Buchel:2013dla, Blaback:2014tfa}. Certain solutions with D6 branes are a bit more subtle in this regard as we now explain.

\subsubsection*{Comments on 6-branes}

Two classes of ``anti-D6''-brane solutions have been investigated in the literature. One class has flat worldvolume \cite{Blaback:2011nz, Blaback:2011pn}, which apart from the singular three-form flux have a naked singularity at large distance such that no global solutions can be found. Perhaps the addition of O6 planes and quantum corrections allows a compact solution for which the large distance problem is absent. We ignore that issue in what follows. If the (anti-)D6 would polarise into a D8 the singularity would be absent, but this polarisation does not occur \cite{Bena:2012tx}. After 3 T-dualities this model, although very crude, captures the essence of the singularity of anti-D3 branes smeared over the tip of the KS throat as explained in \cite{Massai:2012jn}. The D8 `no-polarisation' result \cite{Bena:2012tx} T-dualises into a `no-polarisation' result for D5 branes, which was later verified in \cite{Bena:2012vz}. 

The second class has AdS worldvolume and a compact transversal space as predicted first in \cite{Blaback:2010sj,Blaback:2011nz, Blaback:2011pn} and then studied in much more detail in \cite{Apruzzi:2013yva}. In the latter reference it was shown that, despite the presence of charge dissolved in flux that has the opposite sign to the 6-brane sources, the solutions are SUSY and that furthermore the 3-form singularity is resolved since the D6 branes polarise into D8 branes that sit at a stable position (see also \cite{Gaiotto:2014lca}). 

The different polarisation behavior between these two classes was traced back to the presence of extra terms in the D8 potential that are sensitive to the AdS curvature \cite{Junghans:2014wda}.   We expect similar behavior to be possible for anti-D3 solutions with AdS worldvolume.  This is not in contradiction with the conclusions in this paper since those anti-D3 backgrounds of relevance to KPV \cite{Kachru:2002gs}(and KKLT \cite{Kachru:2003aw}) break SUSY and have a flat worldvolume in the non-compact case (KPV) and de Sitter in the compact case (KKLT)\footnote{To be more precise, in the KKLT scenario on can also lift the SUSY AdS to a non-SUSY AdS with a smaller curvature. But these AdS backgrounds have a parametric hierarchy of scales between the KK scale and the cosmological constant, as opposed to the anti-brane solutions with AdS worldvolume at tree-level.}.

The (anti-)D6 solutions with $AdS_7$ worldvolume are such that a solution without Romans mass is also possible and carries a similar 3-form singularity. Without Romans mass it is less natural to think of D8 polarisation since there is no $F_0$-flux that would cause the D6 to polarise. What saves the day for the 3-form singularity, in the absence of Romans mass, is the lift to 11-dimensional supergravity (which is not possible when $F_0\neq 0$) to the well-known Freund-Rubin vacuum $AdS_7\times S^4$. This implies that the 3-form singularity, as such in 10D, is harmless and one could be tempted to conclude that this implies a diverging $H^2$  is harmless with non-zero Romans mass as well, and by extension the same could be said for anti-D3 solutions. This is incorrect and  the resolution in the massless $AdS_7$ solution does not contradict the findings of this paper. In fact one could argue it strengthens our interpretation, because in the massless case, the 3-form flux does not induce D6 charges\footnote{Induced D6 charge is proportional to $\int F_0 H$. } and there is no possibility for brane-flux annihilation. In that sense it is interesting that the single solution that is known to be stable and has diverging $H^2$, that remains unresolved in 10D, is such that the fluxes do not induce brane charge and hence there is no channel for brane-flux decay and the worries expressed in this paper do not apply.

\subsection{What if it is resolved at small scales?}

If the singularity is resolved beyond SUGRA, this implies a problem of a different kind since it is rather unlikely that it can be verified whether effects, such as an infinite tower of derivative corrections, smoothens the singularity. In practice one could see this as a problem similar, or equal, to the Dine-Seiberg problem \cite{Dine:1985he}. The Dine-Seiberg problem can be stated as the problem that a typical vacuum is uncomputable since it is expected to live a strong coupling or large curvature. It was the original motivation of the flux compactification program to get around the Dine-Seiberg problem using fluxes as a tree-level source of vacuum energy that could create vacua at small coupling and large volume. However, if the KKLT scenario \cite{Kachru:2003aw} can only be verified to work by computing the $H^2$ behavior near the anti-D3 brane, where curvatures and flux gradients are tremendous, then one is back to square one (at least for dS vacua, not for SUSY AdS vacua).

\subsection{Conclusion}
We have elaborated on \cite{Blaback:2012nf}, which provides an interpretation of the 3-form singularities typical to anti-brane solutions. We claim that, in the absence of a resolution of the singularity within supergravity (at large length scales), the solution is necessarily unstable and decays perturbatively to the SUSY vacuum. The meta-stability of the KPV and KKLT vacuum is then lost and the time-dependent process itself is what regularizes the 3-form singularity.  

Apart from the arguments in this paper there might exist hints for an instability from different view points. In particular we have in mind the tachyon found in \cite{Bena:2014jaa}, which is valid inside the Polchinski--Strassler regime (see also \cite{Bena:2014bxa}). A yet different criterium for meta-stable SUSY breaking in holographic backgrounds is suggested in \cite{Argurio:2013uba} and amounts to computing poles in holographic correlation functions to unveil the presence of tachyonic modes.

We are aware of two arguments in favor of the KPV meta-stable state and against the interpretation of \cite{Blaback:2012nf} and this paper. 

First, one can argue that, if the KPV meta-stable state does not exist it would provide the first failure of a probe computation in physics and therefore this is rather unlikely. But as we have mentioned in the introduction, the probe regime implies arbitrary small $p/M$ and it that regime we have no control over the NS5-brane actions since the wrapped $S^2$'s are of stringy size. Also, the probe computations in \cite{Kachru:2002gs} were carried out at strong coupling so it is difficult to say whether the probe computation was valid at the start.

A second argument is closely related to the first and relies on standard lore in effective field theory.  If one perturbes a stable state without flat directions one should not find an instability. The state should be gapped and hence a tunable small parameter cannot make things unstable. One way to understand why this could fail is the presence of a small throat region around the brane. A throat region can indeed redshift the tension of the anti-brane and hence its associated uplift energy, but the masses of local open string moduli, that determine the stability of the anti-brane are red-shifted as well \cite{Aharony:2005ez}. For instance we find that flux gradients, which should be described as KK modes, couple to the open string modes (the NS5 position) and make the set-up unstable. Normally KK modes are heavy, but due to the throat region, they are locally light and help to destabilize the brane.

It is essential to understand that our arguments have little to do with the size of the back-reaction being ``large''. Large would mean that the perturbations to the geometry would extend to the UV of the KS throat. This would imply, in the KKLT scenario that the uplift energy is stringy size and the dS vacuum is lost. This is not what we claim happens\footnote{Such unexpected things do have been shown to happen in other brane constructions used in string cosmology \cite{Conlon:2011qp}. }. In fact the 3-form singularity is integrable in exactly such a way that the uplift term remains as what was guessed in the original KKLT scenario \cite{Junghans:2014xfa}. 

In any case, it is our opinion that the safest conclusion at this point in time is that anti-branes cannot be used to break SUSY in a controllable manner, which is an indication that string theory might be more restrictive at low energies than assumed thus far.

\section*{Acknowledgements}
We are happy to thank Johan Bl{\aa }b\"ack for earlier collaborations and discussions related to this paper. We would also like to thank Riccardo Argurio, Juan Diaz Doronsorro, Daniel Junghans, Stanislav Kuperstein, Stefano Massai, Lubo$\check{s}$ Motl, Brecht Truijen and Alessandro Tomasiello  for useful discussions. We furthermore thank the organisers of the INI workshop ```Supersymmetry breaking in string theory'' and the IFT workshop `` Fine-Tuning, Anthropics and the String Landscape''  for organising  inspiring workshops. UD is supported by the Swedish Research Council (VR). TVR is supported by a Pegasus fellowship and by the Odysseus programme of the FWO Vlaanderen. TVR furthermore acknowledges support from the European Science Foundation Holograv Network.

\appendix

\section{Bubble nucleation rate}

Let us now calculate the tunneling amplitude as we approach the critical value for $\lambda$. To do this we must first evaluate the effective tension of the bubble wall. The Euclidean action can be written
\be
S_E= \int_{\tau_0} ^{\tau_1} d\tau L(\psi )=\int_{\psi_0} ^{\psi_1} d\psi P - \int_{\tau_0} ^{\tau_1} d\tau H(\psi ) ,
\ee
where $\psi_0$ and $\psi_1$ are the turning points such that $\psi_0$ is the meta-stable minimum, while $\psi_1$ is on the slope on the other side of the barrier.  The tension is given by
\be
\sigma = b_0\sqrt{g_sM} \int_{\psi_0} ^{\psi_1} d \psi P
\ee
where $P$ is the Euclidean momentum given by
\be
P^2=m^2-(V-E_0)^2.
\ee
In our case $m$ is the tension of the NS5-brane and $V$ the potential due to the applied $H$-field. The energy $E_0$ is adjusted so that the $\psi_0$ turning point is at the meta-stable minimum.
We define
\be
W=m+V-E_0=V_{\rm eff} -E_0,
\ee
with $V_{\rm eff}$ as before. We then get
\be
P^2=W(2m-W).
\ee
When $P<<m$, the second factor can be approximated by $2m$, and we are under the barrier as soon as $W>0$. In our case, with $V<0$ and $E_0>0$, it is the sign of $W$ that determines the barrier even when $P$ is not small.
  
When we are close to the critical point, we can use the scaling introduced around equation (\ref{potential}), and write 
\be
W=\frac{p}{M}\Bigl( \sqrt{1+x^4}-ax^3\Bigr) -E_0,
\ee
where we for convenience have introduced $
x=b_0  \psi \Bigl(\frac{M}{\pi p} \Bigr) ^{1/2}$ and $a=\frac{2}{3}b_0^{-3} \Bigl(\sqrt{\frac{\pi p}{M}} \Bigr) \lambda $.
The limits of the integration, and the energy $E_0$,  are chosen such that
\begin{equation}
 W(x_1 )=W(x_0 ) =W'(x_0)=0\,.
\end{equation}
We expand around $x=x_0+y$, and get
\be
W\sim \frac{1}{2} W^{(2)} (x_0 ) y^2 +\frac{1}{6} W^{(3)} (x_0 ) y^3.
\ee
At the critical value of $\lambda_c$ we have that $a=a_c=\frac{\sqrt{2}}{3}$ and so close to the critical value we write $a=\tfrac{\sqrt{2}}{3}-\epsilon ^2$.
This gives
\be
W = \frac{p}{M} \Bigl(\sqrt{3} 2^{1/4}\epsilon y^2 -\tfrac{\sqrt{2}}{3} y^3\Bigr)\,, 
\ee
with $y_0=0$ and $y_1=\tfrac{3}{2}\,2^{1/4} \sqrt6 \epsilon$.

We then find $
P^2\simeq 2\sqrt2 \frac{p}{M}W$, and finally
\be
\sigma = \frac{3}{5} 2^{19/8} \sqrt{\pi g_s M} (\frac{p}{M})^{3/2} (1-\frac{\lambda}{\lambda _c})^{5/4} \,.
\ee
This makes it explicit how the tunneling suppression disappears as the critical value of $\lambda$ is approached.

\bibliographystyle{utphys}
\bibliography{BiblioThomas}
\end{document}